\documentclass[11pt]{article}
\usepackage{amssymb,latexsym,amsmath,amsthm,graphicx,subfigure}

\usepackage{color}

\advance \topmargin by -\headheight
\advance \topmargin by -\headsep
\evensidemargin \oddsidemargin
\newcommand{\la}{\lambda}

\newcommand{\R}{\ensuremath{\mathbb{R}}}

\newtheorem{theorem}{Theorem}

\begin{document}
\title{The Picard--Fuchs equations for complete hyperelliptic integrals of even order curves, and the
actions of the generalized Neumann system
\footnote{AMS Subject Classification 14H70, 34M03, 34M56}}
\author{Yuri Fedorov$^1$ and Chara Pantazi$^2$ \\
 Department de Matem\`atica I, Universitat Politecnica de Catalunya, \\
Barcelona, E-08028 Spain \\
e-mails: \footnotesize{Yuri.Fedorov@upc.edu, Chara.Pantazi@upc.edu}}

\maketitle

\abstract{We consider a family of genus 2 hyperelliptic curves of even order and
obtain explicitly the system of 5 linear ODEs for periods of the corresponding
Abelian integrals of first, second, and third kind, as functions of the parameters of the curves.

The system is an extention of the well studied
Picard--Fuchs equations for periods of complete integrals of first and second kind on odd hyperelliptic curves.

The periods we consider are linear combinations of the action variables of several
integrable systems, in particular the generalized Neumann system with polynomial separable potentials.
Thus the solutions of the extended Picard--Fuchs equations can be
used to study various properties of the actions. }

\section{Introduction}
Given a family of elliptic curves ${\cal E} \subset {\mathbb P}^2$ in the Legendre form
$$
 w^2 = (1-z^2) (1-k^2 z^2),
$$
it is known that the complete elliptic integrals of first kind
$$
K(k) = \int_0^1 \frac {dz}{\sqrt {(1-z^2) (1-k^2 z^2)} } , \quad K'(k) = \int_1^{1/k} \frac {dz}{\sqrt {(1-z^2) (1-k^2 z^2)} }
$$
as functions of the modulus $k \in {\mathbb C}$, give 2 independent solutions of the hypergeometric equation of the
 Legendre type
\begin{equation} \label{hyper}
 k(1-k^2) \frac {d^2 y}{d k^2} - (1+k^2) \frac {d y}{d k} +k y =0,
\end{equation}
that is, $K(k)= \frac{\pi}{2} F(\frac 12, \frac 12, 1; k^2)$. The equation has singular points $z_{1,2,3}=-1,0,1$, which means that the solutions $K(k), K'(k)$
are not single-valued: when $k$ performs a loop around $z_i$, these functions transform to a linear combination of $K(k), K'(k)$. That is, the solutions $y(k)$
undergo a monodromy. The monodromy group is isomorphic to the homology group $H_1({\cal E},{\mathbb C})$.

Equivalently, \eqref{hyper} can be rewritten as a system of first order equations for $K(k)$ and the complete integral of the {\it second kind}\footnote{This integral is slightly different from the canonical integral $E(k)$, for this
reason we use the notation $\bar E(k)$.}
$$
\bar E(k) = \int_0^1 \frac {z^2 dz}{\sqrt {(1-z^2) (1-k^2 z^2)} } \, ,
$$
namely,
\begin{equation} \label{1}
 \frac{d K}{dk}= \frac{1}{k (1-k^2)} (k^2 K-\bar E), \quad  \frac{d \bar E}{dk}= \frac{k}{1-k^2} (K-\bar E)
\end{equation}
(see e.g., \cite{Law}).
\medskip

The above description has been generalized to the case of curves of higher genus (see \cite{Sch}).
As an illustration, consider first a family of the genus $g$ hyperelliptic curves of odd order
\begin{equation} \label{odd_curves}
\Gamma_h  =\{ w^2 = (z-a_1)\cdots (z-a_{g+1})( z^g+h_1 z^{g-1} +\cdots+ h_{g-1} z+ h_g ) \}
\end{equation}
with the parameters $h_1,\dots, h_g\in {\mathbb C}$. Here $a_1,\dots,a_{g+1}$ are distinct constants.
For generic values of $h_i$ the curves are 2-fold covering of ${\mathbb C}=\{ z\}$ ramified at $z=a_1,\dots,a_{g+1}$
and $\rho_1,\dots, \rho_g$, the roots of the polynomial $P_g(z)=z^g+h_1 z^{g-1} +\cdots+ h_{g-1} z+ h_g$.

Consider the following canonical basis of $g$ holomorphic  differentials
and $g$ meromorphic differentials of the second kind on $\Gamma$:
$$
\omega_i = \frac{z^{i-1}\, dz}{w}, \quad \omega_{g+i} = \frac{z^{g-1+i}\, dz}{w} , \qquad i=1,\dots,g.
$$
Let $\gamma \in H_1(\Gamma, C)$ be a cycle on $\Gamma$. Then the periods of the above differentials
\begin{equation} \label{pers_4}
J_1 = \oint_\gamma \omega_1, \quad \dots\, , \quad J_{2g} = \oint_\gamma \omega_{2g}
\end{equation}
also become functions of the parameters $h_1, \dots, h_g$ in \eqref{odd_curves} or of the roots $\rho_1,\dots,\rho_g$.

Note that $J_i$ are not single-valued functions of $h_i$: when these parameters vary in such a way that
one of the roots, say $\rho_1$, goes around $a_i$ or $\rho_2,\dots,\rho_g$,
each integral $J_i$ becomes a linear combination of $J_1,\dots, J_{2g}$, i.e., undergoes a monodromy.

Following the classical theory of differential equations, the integrals
$J_i=J_i(h_1,\dots h_g)$ are solutions of a systems of linear ODEs, with $h_i$ being independent variables,
called the {\it Picard--Fuchs} equations (see, e.g., \cite{Ince}):
\begin{equation} \label{PF_general}
 \frac{\partial J}{\partial h_k} = M_k (h)  \, J, \qquad J=(J_1,\dots, J_{2g})^T, \quad  k=1,\dots,g,
\quad M_k \in GL(2g, {\mathbb C}).
\end{equation}
They are natural generalizations of the Legendre equation \eqref{hyper} or \eqref{1}\footnote{
More precisely, the original Picard--Fuchs equations are second order equations obtained by elimination of the periods
of the meromorphic differentials.}.

Due to the monodromy property,  some of the components of $M_k (h)$ have poles when
one of the roots $\rho_i$ coincides with $a_j$ or with the other roots.

Families of hyperelliptic curves $\Gamma_h$ often appear in quadratures of integrable systems of classical mechanics
and mathematical physics, in particular the Neumann system (see below),
whereas certain linear combinations of the integrals $J_i(h)$ give {\it action} variables
${\cal I}_{1}(h), \dots, {\cal I}_{g}(h)$ of the systems.
Knowledge of properties of such functions is important in study of periodic solutions, in quantization,
in applications of the KAM theory to perturbations of the integrable systems.

The integrals $J_i$ are transcendental functions of $h_j$ and, as mentioned in several publications,
instead of computing them numerically, in some cases it is less expensive to integrate
numerically the above Picard--Fuchs equations, at least locally.

Following this idea, the authors of \cite{DV2} derived differential equations for the periods $J_i$
for any genus $g$, taking however, as an independent variable
one of the roots $\rho_i$ in \eqref{odd_curves}, and not a constant of motion $h_k$.
(Thus, they obtained the {\it Gauss--Manin} equations.)

A similar approach was followed in \cite{DV1, DDB} to treat the actions of the Kovalevkaya top and the Jacobi problem
on geodesics on a triaxial ellipsoid.

For another basis of meromorphic differentials on $\Gamma_h$, a similar system of Gauss--Manin equations was obtained
in \cite{Enol_Rich}.

The ony disadvantage of this approach is the dependence of all the constants $h_k$ on any root $\rho_i$, which makes
it difficult to study the properties of $J_i(h)$ as a function of one $h_k$, when all the other ones are fixed.

The choice of $\rho_i$ instead of $h_k$ was motivated in \cite{DV2} by the observation that
the Picard--Fuchs equations with the independent variables $h_k$ become highly cumbersome even for the lowest
non-trivial case $g=2$.
\medskip

The purpose of this note is to derive the Picard--Fuchs equations of type \eqref{PF_general}
for the case of the family of even order genus 2 curves
$$
\Gamma_h=\{ w^2=(z-a_1)(z-a_2)(z-a_3)(z^3+h_1 z+ h_2)\},
$$
which appear in quadratures of an integrable generalization of the Neumann system with a separable quartic potential.

We observe that, in contrast to the odd order curves \eqref{odd_curves} and the equations \eqref{PF_general},
in our case the order of the PF equations is 5, since they also include an Abelian integral of 3rd kind. To our knowledge,
such case was not considered before.

The equations are written in a quite compact and symmetric form, suitable for possible applications.

\section{The classical Neumann system and its generalization}

 Recall that the Neumann system describes the motion of a point on the unit sphere $S^{n-1}=\{\langle x,x \rangle=1\}$,
$x\in\R^n$ under the action of the quadratic potential $U=\langle x,Ax\rangle/2$, $A$ being diagonal matrix with
constant eigenvalues $a_1,a_2,\dots, a_{n}$. The Hamiltonian of the problem has the form
$$
H(x,y)=\frac{1}{2}(|y|^2|x|^2-\langle y,x \rangle^2)+\frac{1}{2} \langle x,Ax\rangle,
$$
where $p\in T_x S^{n-1}$ is the momentum (see e.g., \cite{Moser_var,Moser_Chern}).

Neumann (\cite{Neum}, 1856) considered this problem in the case $n=3$
and solved it completely in terms of theta-functions of 2 variables.

In the elliptic (spheroconical) coordinates $\lambda_1, \dots, \lambda_{n-1}$ on $S^{n-1}$ such that
$$
x_i^2 = \frac{(a_i-{\la}_1)\cdots (a_i-{\la}_{n-1})}
{(a_i-a_1)\cdots(a_i-a_n)} \, , \qquad i=1,\dots, n
$$
and in the corresponding conjugated momenta, the Hamiltonian takes St\"ackel form, and the system is reduced to the quadratures
\begin{gather}
\frac{ \la_1^k d\la_1}{\sqrt{ R(\la_1)}} + \cdots + \frac{\la_{n-1}^k d\la_{n-1}}{\sqrt{R(\la_{n-1})}}
 = \left \{\begin{aligned} & 0 \quad \text{if $k=0,1,\dots, n-2$,} \\
                          & dt \quad \text{if $k=n-1$,} \end{aligned} \right. \qquad k=0,1,\dots, {n-2}  \label{quad2} \\
R (\la) = \Phi(\la) P_{n-1}(\la), \quad
\Phi(\la) = (\la-a_1) \cdots (\la-a_n), \notag \\
P_{n-1}(\la)= \la^{n-1} + h_1\la^{n-2} + \cdots+ h_{n-1}=(\la-\rho_1)\cdots (\la-\rho_{n-1}) , \nonumber
\end{gather}
where $h_1,\dots, h_{n-1}$ are constants of motion. 

Here the differentials $\la^k d\la/ \sqrt{ R(\la)}$ can be regarded as holomorphic differentials on the genus $g=n-1$
hyperelliptic curve $\Gamma_h =\{\mu^2= \Phi(\la) P_{n-1}(\la) \}$, already described in \eqref{odd_curves}.

By integrating the quadratures \eqref{quad2} and inverting the integrals,
symmetric functions of the elliptic coordinates $\la_j$ and, therefore, the
Cartesian coordinates $x_i$, can be expressed in terms of theta-functions of $u_k$ and, therefore, of the time $t$
(see \cite{Mum_Theta}).
The generic real invariant varieties are unions of $n-1$ dimensional tori ${\mathbb T}^{n-1}$. Moreover, the tori are
real parts of complex Abelian varieties, which are isogeneous to the Jacobians of the curves, and the system is
algebraic integrable (see \cite{Moser_var, Mum_Theta}).

On the other hand, as was shown in several publications (see e.g., \cite{Enol, 16}),
the Neumann system admits a hierarchy of integrable
generalizations, in which the quadratic potential $U(x)=\langle x,Ax\rangle/2$
is replaced by polynomial or rational potentials,
which are all separable in the same elliptic coordinates. For all such generalizations, the dimension of the generic
invariant tori is the same, $n-1$. On the other hand, for a class of separable polynomial potentials of degree $2N$,
the quadratures take the following form, which generalizes \eqref{quad2}:
\begin{gather}
\frac{ \la_1^k d\la_1}{\sqrt{ {\cal R} (\la_1)}} + \cdots + \frac{\la_{n-1}^k \, d\la_{n-1}}{\sqrt{{\cal R}(\la_{n-1})}}
 = \left \{\begin{aligned} & 0 \; \text{if $k=0,1,\dots, n-2$,} \\
                          & dt \; \text{if $k=n-1$,} \end{aligned} \right. \qquad k=0,1,\dots, n-2,  \label{quadN}
\end{gather}
where now ${\cal R} (\la) = \Phi(\la) {\cal P}_{N+1}(\la)$,
$$
\Phi(\la) = (\la-a_1) \cdots (\la-a_n), \quad
{\cal P}_{N+1}(\la)= \la^{N+1} + h_1\la^{n-2} + \cdots+ h_{n-1} .
$$ 
The quadratures include $n-1$ holomorphic differentials on the hyperelliptic curve
$$
\Gamma_h =\{\mu^2= \Phi(\la) \, {\cal P}_{N+1}(\la)  \}
$$
of genus $g=[(n+N)/2]$. This implies that for the separable potentials of degree $2N>4$, the genus of $\Gamma_h$
is bigger than the dimension of the tori, and one can show that in this case the system is no more algebraic integrable
(\cite{AF,Vanh}).

The action variables of the original and generalized Neumann systems are the periods of the Abelian integrals
$$
{\mathcal  J}_j (h_1,\dots,h_{n-1}) = \frac{1}{2\pi} \oint_{\gamma_j}
\frac{(\la^{N+1} + h_1\la^{n-2} + \cdots+ h_{n-1})\,  d\la}{\sqrt{{\cal R}(\la)}},  \qquad j=1,\dots, n-1 ,
$$
$\gamma_j$ being certain cycles on the Riemann surface $\Gamma_h$.
Note that the functions ${\mathcal  J}_j (h_1,\dots,h_{g})$
are also the frequencies of the angle variables on the
tori ${\mathbb T}^{n-1}$. Then a solution to the Neumann system is periodic if and only if the quantities
${\mathcal  J}_j$ are {\it commensurable}.
So, knowledge of ${\mathcal  J}_j (h)$ is important in describing periodic solutions of the system.

As follows from the above, the action variables ${\mathcal J}_j$ are linear combinations of the periods of the
following basic $g$ holomorphic and $g$ meromorphic differentials on $\Gamma_h$
\begin{equation} \label{diffs_N}
J_k = \oint_\gamma \omega_k, \qquad \omega_s= \frac{ \la^{s-1} d\la}{\sqrt{ {\cal R} (\la)}}, \quad
\omega_{g+s}= \frac{ \la^{g+ s-1} d\la}{\sqrt{{\cal R}(\la)}} , \quad s=1,\dots,g.
\end{equation}
For the classical Neumann system with the quadratic potential ($N=1$) the above $2g$ differentials
satisfy the Picard--Fuch equations \eqref{PF_general}.
However, for $N>1$ this is not always true. 

For concreteness, below we restrict ourselves to the simplest case $n=3$ and the quartic separable potential ($N=2$)
$$
U(x)= \langle x,Ax \rangle ^2- 2 \text{Tr} A \langle x,Ax \rangle - \langle x,A^*x \rangle, \quad A^*=\det A \, A^{-1}.
$$
In the elliptic coordinates it reads $(\la_1^3-\la_2^3)/(\la_1-\la_2)=\la_1^2+ \la_1 \la_2+\la_2^2$ and,
 due to \eqref{quadN}, corresponds to genus 2 even order curves
\begin{align} \label{g3}
w^2 & =(\la-a_1)(\la-a_2)(\la-a_3)\cdot(\la^3+h_1 \la+ h_2), \quad \text{or} \\
w^2 & =(\la-a_1)(\la-a_2)(\la-a_3)\cdot(\la-\rho_1)(\la-\rho_2)(\la-\rho_3),  \notag
\end{align}
whose compactifications in ${\mathbb P}^2$ have 2 infinite points, which we denote by $\infty_-, \infty_+$.
It follows that
\begin{equation} \label{hs}
h_1= \rho_1 \rho_2+\rho_1\rho_3+\rho_2\rho_3, \quad h_2=-\rho_1\rho_2\rho_3, \quad \text{and also} \quad -\rho_1-\rho_2-\rho_3=h_3=0 .
\end{equation}

The differentials \eqref{diffs_N} then are
$$
\omega_1=\dfrac{d\la}{w}, \quad \omega_2=\dfrac{\la \,d\la}{w}, \quad
\omega_3=\dfrac{\la^2\,d\la}{w}, \quad \omega_4=\dfrac{\la^3\, d\la}{w}.
$$
One observes that, in contrast to $\omega_4$, the differential $\omega_3$ is meromorphic of the 3rd kind, i.e., it has
a pair of simple poles at $\infty_-, \infty_+$, and that the corresponding periods $J_1,\dots, J_4$ do not form
a closed system of differential equations with respect to the constants $h_1$ or $h_2$. It turns out that in this case
the Picard--Fuchs equations must include also the period $J_5$ of the differential of the second kind
$\omega_5=\dfrac{\la^4\,d\la}{w}$. Thus, these equations are of order 5.

\section{The Picard--Fuchs equations for genus 2 even order curves}

To derive the Picard--Fuchs equations for the considered case,
we first compute the derivatives of the integrals $J_1,\dots, J_5$
with respect to the roots $\rho_\alpha$ in \eqref{g3}. Namely, rewrite the genus 2 curve in the form
$$
w^2=R(\la), \quad R(\la)=(\la-e_1)(\la-e_2)(\la-e_3)(\la-e_4)(\la-e_5)(\la-e_6).
$$
Like in several other publications (see, e.g., \cite{Enol}), we will use the following key relation
\begin{gather}
\label{enol}
A_j^{(k)} \frac{\partial }{\partial e_k} \left(\dfrac{\la^j}{w}\right)
=\dfrac{ a_j^{(k)} \la^4+b_j^{(k)}\la^3+c_j^{(k)}\la^2+d_j^{(k)}\la+ g_j^{(k)} }{w}-\dfrac{d}{d\la} \left( \dfrac{w}{\la-e_k} \right) ,\\
j=0,1,\dots,4, \quad k=1,\dots, 6, \notag
\end{gather}
where $A_j^{(k)}, a_j^{(k)}, \dots,g_j^{(k)}$ are functions of the branch points $e_i$ only. Namely, let us write
$$
R'(e_k)= \frac{d R(\la)}{d\la}\bigg|_{\la=e_k} = e_k^5+ \Delta_1^{(k)} e_k^4 + \Delta_2^{(k)} e_k^3 + \Delta_3^{(k)} e_k^2 + \Delta_4^{(k)} e_k
+ \Delta_5^{(k)},
$$
so that the coefficients $\Delta_i^{(k)}$ are elementary symmetric functions of $\{e_1,\dots,e_6\}\setminus e_k$ of degree
$i$. In particular,
$$
 \Delta_1^{(1)}=-e_2-e_3-e_4-e_5-e_6, \quad \Delta_5^{(1)}=-\,e_2 e_3 e_4 e_5 e_6.
$$
Then comparing both sides of \eqref{enol}, we obtain
\begin{align}
A_j^{(k)} & =\dfrac{R'(e_k)}{e_k^j},  \notag \\
a_0^{(k)} & =a_1^{(k)}=a_2^{(k)}=a_3^{(k)}=a_4^{(k)}=A^{(k)} = 2, \notag \\
b_0^{(k)} & =b_1^{(k)}=b_2^{(k)}=b_3^{(k)}= B^{(k)}  = -\dfrac{1}{2}\left(e_k-3\Delta_1^{(k)}\right), \quad
b_4^{(k)}=B + \dfrac{R'(e_k)}{2 e_k^4}, \label{ABC} \\
c_0^{(k)} & =c_1^{(k)}=c_2^{(k)}=C^{(k)} = -\dfrac{1}{2}(e_k^2+ e_k\Delta_1 ^{(k)} -2 \Delta_2^{(k)}) , \notag \\
c_3^{(k)}& =c_4^{(k)}=C^{(k)} + \dfrac{R'(e_k)}{2 e_k^3}, \notag \\
d_0^{(k)} & =d_1^{(k)}=D^{(k)} = -\dfrac{1}{2}\left(e_k^3 + e_k^2\Delta_1^{(k)}+e_k\Delta_2^{(k)} -\Delta_3^{(k)}\right),
\notag \\
d_2^{(k)} & =d_3^{(k)}=d_4^{(k)}=D^{(k)}  + \dfrac{R'(e_k)}{2 e_k^2}, \notag \\
g_0^{(k)}& =G^{(k)} =-\dfrac{1}{2}\left(e_k^4+e_k^3 \Delta_1^{(k)}
+e_k^2 \Delta_2^{(k)} + e_k \Delta_3^{(k)}\right), \notag \\
g_1^{(k)}& =g_2^{(k)}=g_3^{(k)}=g_4^{(k)}= G^{(k)} +\dfrac{R'(e_k)}{2 e_k}. \notag
\end{align}

Multiplying the both sides of \eqref{enol} by $d\la$, and using again the notation
\begin{equation} \label{omegas}
\omega_1=\dfrac{d\la}{w}, \quad \omega_2=\dfrac{\la d\la}{w}, \quad \omega_3=\dfrac{\la^2d\la}{w}, \quad
\omega_4=\dfrac{\la^3d\la}{w}, \quad \omega_5=\dfrac{\la^4d\la}{w},
\end{equation}
one gets
\begin{gather}
\label{enol1}
\frac{\partial}{\partial e_k }\omega_i = \frac{e_k^j}{R'(e_k)} \left( g_j^{(k)} \omega_1 + d_j^{(k)} \omega_2 +
c_j^{(k)} \omega_3+  b_j^{(k)} \omega_4 + a_j^{(k)} \omega_5 + d F_k \right), \\
F_k=\dfrac{w}{\la-e_k}, \quad j=i-1, \quad i=1,\dots, 5.   \notag
\end{gather}
Since
$$
\frac{\partial}{\partial e_k } \left( \oint_\gamma \omega_i \right) =\oint_\gamma \frac{\partial }{\partial e_k} \omega_i ,
$$
and since $dF_k$ is a differential of a meromorphic function of $\Gamma_h$, from \eqref{enol1} we obtain the
following system for the vector of periods $J=(J_1,\dots, J_5)^t$:

\begin{gather}
2 \frac{\partial J}{\partial e_k}= {\cal M}_k J, \qquad k=1, \dots, 6, \label{PFe}  \\
{\cal M}_k = \frac{1}{ R'(e_k)} \begin{pmatrix} 1 \\ e_k \\ e_k^2 \\ e_k^3 \\ e_k^4 \end{pmatrix}
( G^{(k)} \, D^{(k)} \, C^{(k)} \, B^{(k)} \, A^{(k)} \,) +
\begin{pmatrix} 0 & 0 & 0 & 0 & 0 \\
 		    1 & 0 & 0 & 0 & 0 \\
		    e_k & 1 & 0 & 0 & 0 \\
		    e_k^2 & e_k & 1 & 0 & 0 \\
		 e_k^3 & e_k^2 & e_k & 1 & 0
\end{pmatrix} \, ,  \label{mat}
\end{gather}
with $G^{(k)}, D^{(k)}, C^{(k)}, B^{(k)}, A^{(k)}$ defined in \eqref{ABC}.

The structure of the matrices ${\cal M}_k$ is similar to that of the Picard--Fuchs equations obtained in
\cite{DV2, Enol_Rich}, however, not the same: the system \eqref{PFe}, \eqref{mat} has an odd order.

Apparently, there does not exist a linear combination of $\omega_1,\dots, \omega_5$ giving a total differential
on $\Gamma_h$, i.e., there is no linear relation between the periods $J_1,\dots,J_5$, which could be used to
reduce the order of the systems \eqref{PFe}.

Now, taking into account \eqref{g3}, we identify the roots $\rho_1,\rho_2,\rho_3$ with $e_1,e_2,e_3$, and
the parameters $a_1, a_2, a_3$ with $e_4,e_5,e_6$.
In view of \eqref{hs}, the following relation between the partial derivatives holds
$$
\begin{array}{l}
\left(
\begin{array}{c}
\dfrac{\partial J_i}{\partial \rho_1}\\
\\
\dfrac{\partial J_i}{\partial \rho_2}\\
\\
\dfrac{\partial J_i}{\partial \rho_3}\\
\end{array}
\right)=
\left(
\begin{array}{ccc}
1  & \rho_2+\rho_3 & -\rho_2 \rho_3\\
 & & \\
1  & \rho_1+\rho_3 & -\rho_1 \rho_3\\
 &  & \\
1 &\rho_2+\rho_1, &-\rho_1\rho_2\\
\end{array}
\right)
\left(
\begin{array}{c}
\dfrac{\partial J_i}{\partial h_3}\\
\\
\dfrac{\partial J_i}{\partial h_1}\\
\\
\dfrac{\partial J_i}{\partial h_2}\\
\end{array}
\right)
\end{array}, \quad i=1,\dots, 5.
$$
Then
\begin{gather}\label{eqmatrix}
\left(
\begin{array}{c}
\dfrac{\partial J_i}{\partial h_3}\\
\\
\dfrac{\partial J_i}{\partial h_1}\\
\\
\dfrac{\partial J_i}{\partial h_2}\\
\end{array}
\right)=\dfrac{1}{\Delta}
\left(
\begin{array}{ccc}
-\rho_1^2(\rho_2-\rho_3) & \rho_2^2(\rho_1-\rho_3) & -\rho_3^2(\rho_1-\rho_2)\\
  & & \\
\rho_1(\rho_2-\rho_3)& \rho_2(\rho_3-\rho_1)  & \rho_3(\rho_1-\rho_2)\\
&  & \\
\rho_2-\rho_3   & \rho_3-\rho_1 & \rho_1-\rho_2
\end{array}
\right)
\left(
\begin{array}{c}
\dfrac{\partial J_i}{\partial \rho_1}\\
\\
\dfrac{\partial J_i}{\partial \rho_2}\\
\\
\dfrac{\partial J_i}{\partial \rho_3}\\
\end{array}
\right), \\
\Delta=(\rho_1-\rho_2)(\rho_3-\rho_1)(\rho_3-\rho_2). \notag
\end{gather}

Now combining the above relations with the equations \eqref{PFe}, and taking into account \eqref{ABC}, \eqref{hs},
we arrive at

\begin{theorem} The vector of periods $J=(J_1,\dots, J_5)^T$ of the differentials \eqref{omegas}
of the even order curve \eqref{g3} satisfies the equations
\begin{gather}
2 \frac{\partial J}{\partial h_1}= {\cal U}_1 J, \quad
2 \frac{\partial J}{\partial h_2}= {\cal U}_2 J,  \label{PFF}  \\
{\cal U}_1 = \sum_{\alpha=1}^3 \frac{1}{ \Phi(\rho_\alpha) }
\frac{\rho_\alpha }{ ( \rho_\alpha -\rho_\beta)^2 (\rho_\alpha-\rho_\gamma)^2 }
 {\bf S}_\alpha +
\begin{pmatrix} 0 & 0 & 0 & 0 & 0 \\
 		    0 & 0 & 0 & 0 & 0 \\
		    1 & 0 & 0 & 0 & 0 \\
		    h_3 & 1 & 0 & 0 & 0 \\
		    h_1 & h_3 & 1 & 0 & 0
\end{pmatrix} \, ,  \notag \\
{\cal U}_2 = \sum_{\alpha=1}^3 \frac{1}{\Phi(\rho_\alpha) }
\frac{ 1 }{ ( \rho_\alpha -\rho_\beta)^2 (\rho_\alpha-\rho_\gamma)^2 }
{\bf S}_\alpha +
\begin{pmatrix} 0 & 0 & 0 & 0 & 0 \\
 		    0 & 0 & 0 & 0 & 0 \\
		    0 & 0 & 0 & 0 & 0 \\
		    1 & 0 & 0 & 0 & 0 \\
		    h_3 & 1 & 0 & 0 & 0
\end{pmatrix} \, , \notag
\end{gather}
where
\begin{gather*}
\Phi( \rho_\alpha )= (\rho_\alpha-a_1)(\rho_\alpha -a_2)(\rho_\alpha -a_3), \quad (\alpha,\beta,\gamma)= (1,2,3),
\quad h_3=0, \notag \\
{\bf S}_\alpha = \begin{pmatrix} 1 \\ \rho_\alpha \\ \rho_\alpha^2 \\ \rho_\alpha^3 \\ \rho_\alpha^4 \end{pmatrix}
(G^{(\alpha)} \, D^{(\alpha)} \, C^{(\alpha)} \, B^{(\alpha)} \, A^{(\alpha)} \,), \notag \\
A^{(\alpha)} =2, \quad B^{(\alpha)} = -\frac{\rho_\alpha^2}{2}- \frac 32 \rho_\alpha + \frac 32 \sigma_1, \quad
C^{(\alpha)} = h_1 - \frac 12 \sigma_1 \rho_\alpha +\sigma_2,  \notag \\
D^{(\alpha)} = \frac 12 \left(-2 \rho_\alpha^3+ \rho_\alpha^2 \sigma_1 + h_2-h_1 \sigma_1-\sigma_3\right), \quad \notag
G^{(\alpha)} = -\rho_\alpha \Phi(\rho_\alpha) + \frac 12 \left( h_2 \rho_\alpha-\sigma_3 \rho_\alpha-h_2 \sigma_1 \right) ,
\end{gather*}
and
$\sigma_1=a_1+a_2+a_3,\; \sigma_2=a_1 a_2+a_3 a_1+a_2 a_3$, $\sigma_3=a_1 a_2 a_3$.
\end{theorem}

The proof is direct and uses the identities
$$
\rho_1^k(\rho_2-\rho_3)+ \rho_2^k(\rho_3-\rho_1)+ \rho_3^k(\rho_1-\rho_2)=\left\{ \begin{aligned} 0, & \quad k=1 \\
									-(\rho_1-\rho_2)(\rho_3-\rho_1)(\rho_3-\rho_2), & \quad k=2 \\
									 (\rho_1-\rho_2)(\rho_3-\rho_1)(\rho_3-\rho_2) h_3, & \quad k=3 \\
						 		  (\rho_1-\rho_2)(\rho_3-\rho_1)(\rho_3-\rho_2) (h_3^2-h_1), & \quad k=4.
\end{aligned} \right.
$$


\section*{Acknowledgements}
The authors thank Prof. V. Enolski for the discussion and usuful suggestions. \\
Both authors are supported by the MICIIN/FEDER grant number
MTM2009-06973. C.P. is also partially supported by the
MICIIN/FEDER grant MTM2008--03437 and by the Generalitat de
Catalunya grant number 2009SGR859.
\footnote{$^2$ Departament de Matem\`atica Aplicada I, Universitat Polit\`ecnica de Cata\-lunya, (EPSEB), Av. Doctor Mara\~{n}\'on, 44--50, 08028 Barcelona, Spain}

\end{document}